\documentclass[prl, superscriptaddress, showpacs, floatfix,
twocolumn]{revtex4}

\usepackage[]{graphicx}
\usepackage{amsmath}
\usepackage{amsfonts}

\def\one{{\mathchoice{\rm 1\mskip-4mu l}{\rm 1\mskip-4mu l}{\rm 1\mskip-4.5mu
l}{\rm
1\mskip-5mu l}}}
\def\ket#1{| #1 \rangle}
\def\bra#1{\langle #1 }
\def\kb#1#2{|#1\rangle\!\langle #2 |}

\def\B{{\cal B}}
\def\C{{\cal C}}

\def\E{{\cal E}}
\def\H{{\cal H}}

\def\P{{\mathcal P}}
\def\R{{\cal R}}

\def\V{{\cal V}}

\newcommand{\qiff}{\quad\text{if and only if}\quad}

\newcommand{\bbC}{{\mathbb{C}}}
\newcommand{\spn}{\operatorname{span}}
\newtheorem{theorem}{Theorem}
\newtheorem{definition}{Definition}

\newtheorem{proposition}{Proposition}
\newtheorem{remark}{Remark}


\begin{document}

\title{Quantum Error Correcting Codes From The Compression Formalism}

\author{Man-Duen~Choi}
\affiliation{Department of Mathematics, University of Toronto, ON
Canada, M5S 2E4}

\author{David~W.~Kribs}
\affiliation{Department of Mathematics and Statistics, University
of Guelph, Guelph, ON, Canada, N1G 2W1} \affiliation{Institute for
Quantum Computing, University of Waterloo, ON Canada, N2L 3G1}

\author{Karol {\.Z}yczkowski}
\affiliation{Perimeter Institute for Theoretical Physics, 31
Caroline St. North, Waterloo, ON, CANADA N2L 2Y5}
\affiliation{Institute of Physics, Jagiellonian University,
       ul. Reymonta 4, {30-059} Cracow, Poland}
\affiliation{Center for Theoretical Physics, Polish Academy of
Sciences, Al.~Lotnik{\'o}w 32/44, {02-668} Warsaw, Poland}

\begin{abstract}
We solve the fundamental quantum error correction problem for
bi-unitary channels on two-qubit Hilbert space. By solving an
algebraic compression problem, we construct qubit codes for such
channels on arbitrary dimension Hilbert space, and identify
correctable codes for Pauli-error models not obtained by the
stabilizer formalism. This is accomplished through an application
of a new tool for error correction in quantum computing called the
``higher-rank numerical range''. We describe its basic properties
and discuss possible further applications.
\end{abstract}

\maketitle


\noindent{\it Introduction.} --- Quantum computers will rely on a
smorgasbord of error correction techniques to combat  harmful
effects such as decoherence on bits of information that are
physically encoded  within quantum systems. The quantum error
correction ``toolbox'' now includes several strategies for
accomplishing these feats, but there are still many deep issues to
resolve. The standard method for error correction via active
intervention in the quantum computing regime has been cleanly
phrased in terms of an analysis of operators on Hilbert space. For
every quantum channel that arises through the usual
system-environment formalism, there is a family of error (or
``Kraus'') operators that describe the possible corruption by the
channel of qubits encoded as states in, or operators on, the
system Hilbert space. The main protocol for quantum error
correction (QEC) \cite{Sho95a,Ste96a,BDSW96a,KL97a} depends upon
the existence and identification of states and operators on which
the error operators are jointly well-behaved in a precise sense.

The stabilizer formalism for QEC \cite{Got96,Got97} gives a
constructive framework to find correctable codes for error models
of ``Pauli type''. While there are other successful techniques
that can be applied in special cases (for instance, see
\cite{CRSS97,RHSS97,ABO99,Rai99b,NC00,KlRo02a,PoRu04,Got05}), the
landscape of general strategies to find codes for other classes of
channels is fairly sparse. In particular, the theory lacks a
systematic method that applies to arbitrary quantum channels.
Indeed, after spending any time at all on this problem, it becomes
clear that this is an extremely daunting challenge.

Nevertheless, in this paper, we introduce an approach based on a
``compression formalism'' that may lead to such a general method.
In particular, we cast the general problem  of finding correctable
codes for quantum channels into a matrix analysis framework. We
then utilize a new tool recently introduced in \cite{CKZ05a} --
called the ``higher-rank numerical range'' -- the study of which
was primarily motivated by this problem. As an application, we
solve the quantum error correction problem in its entirety for the
class of ``bi-unitary channels'' on two-qubit (i.e.,
four-dimensional) Hilbert space, and construct qubit codes for
such channels in the arbitrary dimension case. In the case of
Pauli-error models, we show that this approach captures codes not
obtained via the Pauli matrix stabilizer formalism.

The rest of the paper is organized as follows. Next we recall
basic properties of quantum channels and the formulation of the
active quantum error correction protocol. We follow this by
introducing the higher-rank numerical range as a tool in quantum
error correction. We then consider randomized unitary channels,
and focus on the bi-unitary case. This is followed by a
characterization of correctable codes in the two-qubit case, and a
consideration of the connection with the stabilizer formalism. We
discuss possible further applications and limitations of this
approach throughout the paper, and finish with a conclusion.

\vspace{0.1in}

\noindent{\it Quantum Channels and Active Error Correction.}
--- Consider an open quantum system $S$ represented on a Hilbert
space $\H$, and write $\B(\H)$ for the set of operators that act
on $\H$. A ``snapshot'' of a Hamiltonian-induced evolution of $S$
is called a \textit{quantum channel}. Mathematically, channels are
represented by completely positive, trace preserving maps $\E$ on
$\B(\H)$. (For experimental reasons, the current focus in quantum
computing is on finite-dimensional Hilbert spaces, and thus we
shall make this assumption throughout the paper.) The structure
theorem \cite{Cho75,Kra71} for completely positive maps shows that
every quantum channel $\E$ on $\H$ has an operator-sum
representation of the form $\E(\sigma) = \sum_a E_a \sigma
E_a^\dagger$ for all $\sigma\in\B(\H)$, where the ``error''
operators (or ``Kraus'' operators) $E_a$ are operators that act on
$\H$. A set of error operators  $\{E_a\}$ can always be chosen
with cardinality at most $N^2:=\dim(\H)^2$. For simplicity, we
shall always assume the maximum cardinality holds, by possibly
including zero operators as some of the $E_a$. The trace
preservation condition is equivalent to $\sum_a E_a^\dagger E_a =
\one$. As a notational convenience, we shall write $\E = \{E_a\}$
when the $E_a$ determine $\E$ through the operator-sum
representation, and, as a further convenience, we will also use
this notation when scalar multiples of the $E_a$ are error
operators for $\E$.

In the context of quantum computing, the operators $E_a$ are the
errors induced by the channel $\E$. Thus, error correction
protocols in quantum computing are crafted primarily to mitigate
the effects of such operators on quantum information encoded in
evolving systems. By ``active quantum error correction'', we mean
protocols that involve active intervention into the system to
correct errors. The basic method for active quantum error
correction \cite{Sho95a,Ste96a,BDSW96a,KL97a} identifies {\it
quantum codes} with subspaces $\C$ of the system Hilbert space
$\H$. Then a code $\C$ is correctable for a channel $\E$ if all
states encoded in $\C$ prior to the action of $\E$ can be fully
recovered in a manner allowed by quantum mechanics. From the
operator perspective, {\it $\C$ is correctable for $\E$} if there
is a quantum channel $\R$ on $\H$ such that
\begin{eqnarray}\label{correctdefn}
\big(\R\circ\E \big) (\sigma) = \sigma,
\end{eqnarray}
for all operators $\sigma$ supported on $\C$; that is, all
$\sigma$ of the form $\sigma = P_\C \sigma P_\C$ where $P_\C$ is
the projection of $\H$ onto $\C$.

There is a very useful characterization \cite{BDSW96a,KL97a} of
correctable codes for a given quantum channel $\E$, when a set of
error operators $\E = \{E_a\}$ is known. The code $\C$ is
correctable for $\E=\{E_a\}$ if and only if there is a scalar
matrix $\Lambda = (\lambda_{ab})$ such that
\begin{eqnarray}\label{correctcondition}
P_\C E_a^\dagger E_b P_\C = \lambda_{ab} P_\C \quad \forall a,b.
\end{eqnarray}
Thus, Eq.~(\ref{correctcondition}) shows how the physical problem
of active quantum error correction can be phrased  in terms of a
clean mathematical statement that involves relations satisfied by
the error operators. The prototypical scenario occurs when $\C$ is
correctable and the matrix $\Lambda = \Lambda_C$ is diagonal. Here
the error operators take the code space to mutually orthogonal
subspaces $E_a\C$. In the case that each of the errors restricted
to $\C$ is a scalar multiple of a unitary $U_a$, the correction
operation $\R$ is given by $\R = \{ U_a^\dagger P_a\}$, where
$P_a$ is the projection of $\H$ onto the subspace $U_a\C$.

Let us note that every matrix $\Lambda_C$ which satisfies
Eq.~(\ref{correctcondition}) for some code space $\C$ is
necessarily positive. Indeed, define $E$ to be the $N^2\times N^2$
positive block matrix
\begin{eqnarray}\label{Edefn}
E &=& \left[\begin{matrix} E_1^\dagger \\ E_2^\dagger \\ \vdots
\end{matrix}\right] \left[\begin{matrix} E_1 & E_2 & \cdots \end{matrix}\right]
\geq 0 ,
\end{eqnarray}
where $E_1,E_2,\ldots $ is (any) enumeration of the set $\{E_a\}$,
and $[E_1 \, E_2 \, \cdots ]$ is the operator row matrix mapping
from $\H^{(N^2)}$ to $\H$. (So $E_i^\dagger E_j$ is the $(i,j)$
entry of $E$.) Then observe that the set of equations from
Eq.~(\ref{correctcondition}) can be succinctly stated as the
single matrix equation
\begin{eqnarray}\label{positivity}
0\leq \tilde{P}_\C E \tilde{P}_\C = \Lambda_C \otimes P_\C,
\end{eqnarray}
where $\tilde{P}_\C$ is the $N^2\times N^2$ diagonal block matrix
with $P_\C$ as the operator in each of the diagonal entries. In
fact $\Lambda_C$ is a density matrix when the $E_a$ satisfy the
trace preservation constraint (and $\C$ is non-zero);
\begin{eqnarray}
\big(\sum_a \lambda_{aa}\big) P_\C &=& \sum_a P_\C E_a^\dagger E_a P_\C \\
&=& P_\C \big( \sum_a E_a^\dagger E_a \big) P_\C = P_\C.
\end{eqnarray}

\vspace{0.1in}

\noindent{\it Higher-Rank Numerical Range and Projections.} ---
The characterization from Eq.~(\ref{correctcondition}) of
correctable codes motivates consideration of the following notion.
Let $\sigma$ belong to $\B(\H)$. For $k\geq 1$, define the {\it
rank-$k$ numerical range} of $\sigma$ to be the subset of the
complex plane given by
\begin{eqnarray}\label{nrdefn}
\Lambda_k(\sigma) = \big\{ \lambda\in\bbC : P\sigma P = \lambda P
\,\, {\rm for \,\, some \,\,} P\in\P_k \big\},
\end{eqnarray}
where $\P_k$ is the set of all rank-$k$ projections on $\H$. We
refer to elements of $\Lambda_k(\sigma)$ as ``compression-values''
for $\sigma$, since they are obtained through compressions of
$\sigma$ to $k$-dimensional subspaces. The case $k=1$ yields the
familiar numerical range $W(\sigma)$ for operators \cite{Hal67};
\begin{eqnarray}
\Lambda_1(\sigma) = W(\sigma) = \{ \bra{\sigma\psi}\ket{\psi} \,
:\, \ket{\psi}\in\H,\, ||\,\ket{\psi}\,|| = 1 \}.
\end{eqnarray}
It is clear that
\begin{eqnarray}\label{inclusions}
\Lambda_1(\sigma) \supseteq \Lambda_2(\sigma) \supseteq \ldots
\supseteq \Lambda_N(\sigma).
\end{eqnarray}

Of course, the cases $k>1$ are of immediate interest in quantum
error correction. We refer to the sets $\Lambda_k(\sigma)$, $k>1$,
as the {\it higher-rank numerical ranges}. This notion was
recently introduced in \cite{CKZ05a}, where a number of
mathematical properties were developed. Here we briefly outline
the main points. The following facts from \cite{CKZ05a} apply to
arbitrary $\sigma\in\B(\H)$.

\begin{proposition}\label{genericfacts}
Let $\sigma\in\B(\bbC^N)$. For all $k\geq 1$, the rank-$k$
numerical range $\Lambda_k(\sigma)$ is a compact subset of the
complex plane $\bbC$. If $2k> N$, then $\Lambda_k(\sigma)$ is
either empty, or a singleton set. If $\Lambda_k(\sigma) =
\{\lambda_0\}$ is a singleton set with $2k
> N$, then $\lambda_0$ is an eigenvalue of geometric
multiplicity at least $2k-N$. In particular, $\Lambda_N(\sigma)$
is non-empty if and only if $\sigma$ is a scalar matrix.
\end{proposition}

The main result of \cite{CKZ05a} is stated as follows.

\begin{theorem}\label{Hermitiancase}
If $\sigma=\sigma^\dagger$ is a Hermitian operator on $\H=\bbC^N$,
with eigenvalues (counting multiplicities) given by
\begin{eqnarray}
a_1\leq a_2\leq \ldots \leq a_N,
\end{eqnarray}
and $k\geq 1$ is a fixed positive integer, then the rank-$k$
numerical range $\Lambda_k(\sigma)$ coincides with the set $[a_k,
a_{N-k+1}]$ which is:
\begin{itemize}
\item[$(i)$] a non-degenerate closed interval if  $a_k <
a_{N-k+1}$ , \item[$(ii)$] a singleton set if $a_k =a_{N-k+1}$,
\item[$(iii)$] an empty set if $a_k > a_{N-k+1}$ .
\end{itemize}
\end{theorem}

The cases of non-degenerate spectra for $N=4$ and $N=6$ are
depicted in Figure~1. We note that the proof from \cite{CKZ05a} of
this result is constructive, in the sense that the projections $P$
associated with each $\lambda\in\Lambda_k(\sigma)$ can be
explicitly identified. We shall extend this fact for special cases
in the analysis of the next sections.

\begin{figure} [htbp]
      \begin{center} \
    \includegraphics[width=7.5cm,angle=0]{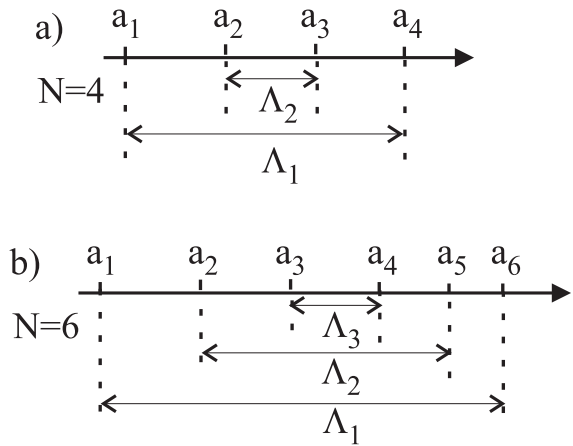}
\caption{Numerical range $\Lambda_k(\sigma)$
   for a non-degenerate Hermitian operator $\sigma$
   of size a) $N=4$ and b) $N=6$
   with spectrum $\{ a_i \}$.}
\label{fig1}
\end{center}
    \end{figure}

Let us consider the following simple examples for the purpose of
illumination.
\begin{itemize}
\item[$(1)$] Let $\sigma=\sigma^\dagger$ be the operator on
2-qubit space $\H = \bbC^2\otimes\bbC^2$ given by
\[
\sigma = \kb{00}{00} + 2 \kb{01}{01} + 3 \kb{10}{10} + 4
\kb{11}{11}.
\]
Then $\Lambda_1(\sigma) = [1,4]$, $\Lambda_2(\sigma) = [2,3]$, and
$\Lambda_3(\sigma)$ and  $\Lambda_4(\sigma)$ are empty.
\item[$(2)$] Consider the Pauli matrix $Z = \kb{0}{0} -
\kb{1}{1}$. Then $\Lambda_1(Z) = [-1,1]$ and $\Lambda_2(Z)$ is
empty. On $\H=\bbC^2\otimes\bbC^2$, if we let $Z_1 =
Z\otimes\one_2$, then $\Lambda_1(Z_1)=[-1,1]$, $\Lambda_2(Z_1) =
[-1,1]$, and $\Lambda_3(Z_1)$ and $\Lambda_4(Z_1)$ are empty. More
generally, if $Z_1 = Z\otimes \one_{2^{n-1}}$ acts on $\H =
(\bbC^2)^{\otimes n}$, then $\Lambda_k(Z_1) = [-1,1]$ for $k\leq
2^{n-1}$.
\end{itemize}

We have extended Theorem~\ref{Hermitiancase} in \cite{CKZ05a}
partly to the case of normal operators; i.e., $\sigma\in\B(\H)$
such that $\sigma\sigma^\dagger = \sigma^\dagger \sigma$. Recall
that the convex hull ${\rm co}\,\Gamma$ of a subset
$\Gamma\subseteq\bbC$ is given by
\begin{eqnarray*}
{\rm co}\,\Gamma = \Big\{ t_1\lambda_1 + \ldots +t_m\lambda_m &:&
 \sum_{j=1}^m t_j = 1,  t_j\geq 0, \\  & &  \lambda_j\in\Gamma, m\geq 1\Big\}.
\end{eqnarray*}
Notice that the following result applies to unitary operators. We
have conjectured in \cite{CKZ05a} that the converse inclusion of
Eq.~(\ref{normalid}) below holds for arbitrary normal operators
$\sigma$.

\begin{theorem}\label{normalcase}
If $\sigma$ is a normal operator on $\H = \bbC^N$, then for all
$k\geq 1$, the rank-$k$ numerical range $\Lambda_k(\sigma)$ is a
subset of every convex hull ${\rm co}\, \Gamma$ where $\Gamma$ is
an $(N+1-k)$-point subset (counting multiplicities) of the
spectrum of $\sigma$; that is,
\begin{eqnarray}\label{normalid}
\Lambda_k(\sigma) \subseteq \bigcap_\Gamma\,\, ({\rm co}\,
\Gamma).
\end{eqnarray}
\end{theorem}

In the next section we shall present a solution to the error
correction problem for a special class of channels. The solution
depends on the analysis of higher-rank numerical ranges discussed
above and specific properties of the channels. Before continuing,
however, let us briefly describe a general method to find
correctable codes based on this tool that may lead to more general
applications. Let $\E = \{E_a\}$ be an arbitrary quantum channel
on $\B(\H)$. Then all correctable codes for $\E$ may be obtained
by applying the following procedure:
\begin{itemize}
\item[$(i)$] For all $a,b$, find scalars $\lambda$ such that
$PE_a^\dagger E_b P = \lambda P$ for some projection $P$.
\item[$(ii)$] For all $\lambda$ from $(i)$, find the projections
$P$. \item[$(iii)$] The set of intersections of ranges of
projections $P_{ab}$ over all distinct pairs $a,b$ from $(ii)$,
corresponds precisely to the set of correctable codes for $\E$.
\end{itemize}

We emphasize that in QEC, the projections $P$ from $(ii)$ must be
explicitly identified for practical applications in quantum
computing. The idea in $(iii)$ will be expanded upon in the
discussion below. We have stated the problem in this form, because
it lends itself to consideration as a ``compression problem''. Of
course, this process is still somewhat abstract. In particular,
step~$(ii)$ will typically require taking infinitely many
intersections. Nevertheless, there may be a way to avoid these
infinities. Let us describe one simplification of the process.
Namely, while the family of operators $E_a^\dagger E_b$,
$\forall\, a,b$, will not be Hermitian in general, the solving of
$(i)$ and $(ii)$ can be reduced to the Hermitian case, the case
for which we have the most general information.

\begin{proposition}\label{Hermitiantreduction}
The projection $P$ is a solution to Eq.~(\ref{correctcondition})
for the family $\{E_a^\dagger E_b:a,b\}$ if and only if $P$ is a
solution to Eq.~(\ref{correctcondition}) for the family of
Hermitian operators $\{E_a^\dagger E_a, T_{ab}^+, T_{ab}^-:
a,b\}$, where
\begin{eqnarray}
T_{ab}^+ &=& E_a^\dagger E_b + E_b^\dagger E_a
\\ T_{ab}^- &=&  i(E_a^\dagger E_b - E_b^\dagger E_a).
\end{eqnarray}
\end{proposition}

{\noindent}{\it Proof.} This simply follows from the fact that the
operator subspace spanned by an operator $A$ and its adjoint
$A^\dagger$, is also spanned by the Hermitian operators
$A+A^\dagger$ and $i(A-A^\dagger)$.  \hfill$\square$

\newpage


\noindent{\it Randomized Unitary Channels.} --- The class of
``randomized unitary channels'' \cite{AL87} form a large class of
physical quantum operations. Such a channel has an operator-sum
representation of the form $\E(\sigma) = \sum_k p_k U_k\sigma
U_k^\dagger$, where the $U_k$ are unitary and the $\{ p_k \}$
determine a probability distribution. The error models $\E =
\{U_k\}$ commonly arise in quantum error correction, quantum
communication and cryptography, quantum information processing and
theory, etc, and are of great physical relevance.

\begin{definition}
A {\bf bi-unitary channel} (BUC) is a randomized unitary channel
$\E=\{V,W\}$ on a Hilbert space $\H$ with an operator-sum
representation consisting of two unitaries; so
\begin{eqnarray}\label{bucdefn}
\E (\sigma) = p V\sigma V^\dagger  + (1-p) W\sigma W^\dagger,
\quad \forall \sigma\in\B(\H),
\end{eqnarray}
for a fixed $p$ with $0< p < 1$.
\end{definition}

We identify correctable qubit codes for all such channels, and we
solve the error correction problem for BUCs on four-dimensional
Hilbert space. The approach we use is constructive in nature and
allows for an explicit identification of the correctable codes. We
shall use properties of the higher-rank numerical range discussed
in the previous section.

There are four non-zero equations to consider in
Eq.~(\ref{positivity}) here. In terms of the matrix $E$, we must
solve the following matrix equation for projections $P$ and
matrices $\Lambda=(\lambda_{ij})$:
\begin{eqnarray}
\left(\begin{matrix} p P & qr PV^\dagger W P
\\ qr PW^\dagger VP & (1-p)P
\end{matrix}\right)  = \left(\begin{matrix} \lambda_{11}P & \lambda_{12}P \\
\lambda_{21}P & \lambda_{22}P
\end{matrix}\right),
\end{eqnarray}
where we have written $q=\sqrt{p}$ and $r=\sqrt{1-p}$. For any
projection $P$, the $\lambda_{11}$ and $\lambda_{22}$ equations
are trivially satisfied with $\lambda_{11} = p$ and $\lambda_{22}
= 1-p$. Further, the $\lambda_{12}$ equation is satisfied if and
only if the $\lambda_{21}$ equation is satisfied. Specifically,
\begin{eqnarray}
qr P V^\dagger W P = \lambda P \qiff qr P W^\dagger V P =
\overline{\lambda} P.
\end{eqnarray}
Thus, we can reduce the entire problem to solving a single
(normalized) equation of the form
\begin{eqnarray}\label{unitaryeqn}
PUP = \lambda P,
\end{eqnarray}
for $\lambda$ and $P$, where $U$ is a unitary on $\H$.

This can also be seen by simply noting that a code is correctable
for the channel $\E = \{ V,W \}$ precisely when it is correctable
for $\E^\prime = \{ \one, V^\dagger W\}$. In fact, for ease of
presentation, when it is convenient we shall assume the channel
$\E$ has the latter form; i.e., there is a unitary $U$ on $\H$
such that
\begin{eqnarray}\label{simpleform}
\E (\sigma) = p \sigma + (1-p) U\sigma U^\dagger, \quad \forall
\sigma \in \B(\H).
\end{eqnarray}

Thus, in the case that $\H=\bbC^4$, we must solve
Eq.~(\ref{unitaryeqn}) for an arbitrary $4\times 4$ unitary $U$.
This follows from the next result.

\begin{theorem}\label{evalues2qubit}
Let $U$ be a unitary on $\H=\bbC^4$. Then we have the following
characterizations of the numerical ranges for $U$:
\begin{itemize}
\item[$(i)$] $\Lambda_1(U)=W(U)$ is the subset of the unit disk in
$\bbC$ given by the convex hull of the eigenvalues for $U$.
\item[$(ii)$] $\Lambda_2(U)$ is non-empty and given as follows.
Let $z_k = e^{i\theta_k}$, with $\theta_k\in [0,2\pi)$ for
$k=1,2,3,4$, be the eigenvalues for $U$, ordered so that $0\leq
\theta_1\leq \theta_2\leq \theta_3 \leq \theta_4<2\pi$.
\begin{itemize}
    \item[$(a)$] If the spectrum of $U$ is non-degenerate, so
$\theta_k\neq \theta_j$ $\forall k\neq j$, then $\Lambda_2(U) =
\{\lambda\}$ where $\lambda$ is the intersection point in $\bbC$
of the line $\ell_{13}$ through $z_1$ and $z_3$, with the line
$\ell_{24}$ through $z_2$ and $z_4$.
   \item[$(b)$] If $U$ has three
distinct eigenvalues, say $\theta_j=\theta_k$ for some pair $j\neq
k$ but $\theta_j \neq \theta_l$ otherwise, then $\Lambda_2(U) =
\{z_j\}$.
   \item[$(c)$] If $U$ has two distinct eigenvalues, each
of multiplicity two, say $z$ and $w$, then $\Lambda_2(U)$ consists
of the line segment
$L = [z,w]$ joining $z$ and $w$.
 \item[$(d)$] If $U$ has two distinct eigenvalues, one $\lambda =
z$ with multiplicity three and the other with multiplicity one,
then $\Lambda_2(U)=\{z\}$.
  \item[$(e)$] If $U$ has a single
eigenvalue $\lambda =z$, then $U$ is the scalar operator $U = z
\one_4$ and $\Lambda_2(U) = \{z\}$.
\end{itemize}
\item[$(iii)$] $\Lambda_3(U)$ is non-empty if and only if
$\Lambda_3(U)=\{\lambda_0\}$ is a singleton set and $\lambda_0$ is
an eigenvalue for $U$ of geometric multiplicity at least three;
\begin{eqnarray}
\dim ( \ker (U-\lambda_0\one)) \geq 3.
\end{eqnarray}
\item[$(iv)$] $\Lambda_4(U)$ is non-empty if and only if $U$ is a
scalar multiple of the identity operator.
\end{itemize}
\end{theorem}

{\noindent}{\it Proof.} The structure of the standard numerical
range $\Lambda_1(U)=W(U)$ follows from well-known matrix analysis
theory \cite{Hal67}, and the structures of $\Lambda_k(U)$,
$k=3,4$, follow from Proposition~\ref{genericfacts} and
Theorem~\ref{normalcase}. The case of interest is that of $(ii)$.

To see $(ii)$, notice that Theorem~\ref{normalcase} can be applied
to show that $\Lambda_2(U)$ is contained in the claimed set for
each of the subcases. (See Fig~2 for a depiction of the four
non-trivial cases $(a)$, $(b)$, $(c)$, and $(d)$.) For the
converse inclusion in each case, we offer a constructive proof of
the required rank-2 projections.

To verify case $(a)$, it suffices to show that the intersection
point $\lambda$ of $\ell_{13}$ and $\ell_{24}$ belongs to
$\Lambda_2(U)$. First solve the following equations,
\begin{eqnarray}\label{rank2scalar}
\left\{ \begin{array}{ccl} \lambda &=& a z_1 + b z_3 \\
\lambda &=& c z_2 + d z_4 \end{array}\right.,
\end{eqnarray}
for nonnegative scalars $a,b,c,d \geq 0$ such that $a + b = 1$ and
$c + d =1$. For instance, $a$ and $b = 1-a$ may be obtained via
the equation
\begin{eqnarray}\label{cosine}
a = \cos^2\theta_a = \frac{\lambda - z_1}{z_1-z_3},
\end{eqnarray}
for some angle $\theta_a$. Then define an orthonormal pair of
vectors $\{\ket{\phi_1},\ket{\phi_2}\}$ by
\begin{eqnarray}\label{rank2vectors}
\left\{ \begin{array}{ccl} \ket{\phi_1} &=& \cos\theta_a
\ket{\psi_1} +
\sin\theta_a \ket{\psi_3} \\
\ket{\phi_2} &=& \cos\theta_c \ket{\psi_2} + \sin\theta_c
\ket{\psi_4} \end{array}\right..
\end{eqnarray}
Then we define a rank-2 projection
\begin{eqnarray}
P = \kb{\phi_1}{\phi_1} + \kb{\phi_2}{\phi_2}.
\end{eqnarray}
Observe that
\begin{eqnarray}
\bra{U\phi_1}\ket{\phi_1} &=& \cos\theta_a z_1
\bra{\psi_1}\ket{\phi_1} + \sin\theta_a z_3
\bra{\psi_3}\ket{\phi_1} \\ &=& a z_1 + b z_3 = \lambda.
\end{eqnarray}
Similarly, we have $\bra{U\phi_2}\ket{\phi_2} = \lambda$. Further,
we also have
\begin{eqnarray}
\bra{U\phi_1}\ket{\phi_2} = 0 = \bra{U\phi_2}\ket{\phi_1}.
\end{eqnarray}
It follows that $PUP = \lambda P$, and hence $\lambda$ belongs to
$\Lambda_2(U)$ as claimed. This verifies case $(a)$.

In case $(b)$, the rank-2 projection $P = \kb{\psi_j}{\psi_j} +
\kb{\psi_k}{\psi_k}$ can be seen to satisfy $ PUP  = z_j P. $ For
case $(d)$, without loss of generality assume $z_1=z_2=z_3=z\neq
z_4$. Then $P=\sum_{j=1}^3 \kb{\psi_j}{\psi_j}$ satisfies $PUP=
zP$. This verifies both case $(b)$ and $(d)$.

Finally, for case $(c)$, we have say $z_1=z_2=z$ and $z_3=z_4=w$.
Let $\lambda$ belong to the line segment $\ell=[w,z]$, and, as
above, solve the following equations for $a,b,c,d$;
\begin{eqnarray}
\left\{ \begin{array}{cclcl} \lambda &=& a z_1 + b z_3 &=& a z + b w \\
\lambda &=& c z_2 + d z_4 &=& c z + d w
\end{array}\right..
\end{eqnarray}
Then define $\{\ket{\phi_1},\ket{\phi_2}\}$ precisely as in
Eqs.~(\ref{cosine},\ref{rank2vectors}). With
$P=\kb{\phi_1}{\phi_1} + \kb{\phi_2}{\phi_2}$, we have
$PUP=\lambda P$. Thus, $\Lambda_2(U)$ coincides with $L=[w,z]$ in
this case, and the result follows. \hfill$\square$

\begin{figure} [htbp]
      \begin{center} \
  \includegraphics[width=8.2cm,angle=0]{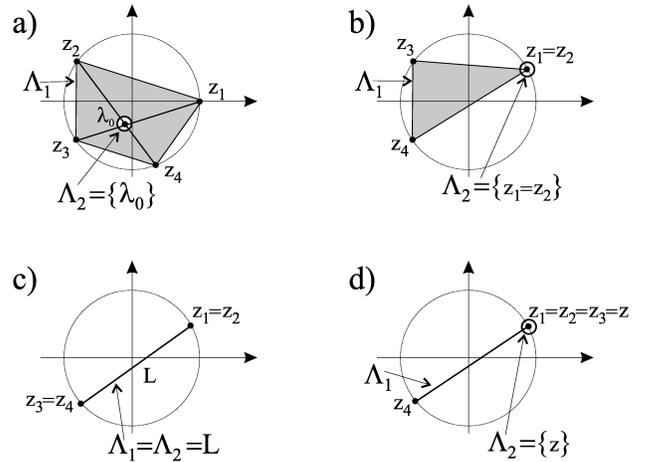}
\caption{Numerical ranges $\Lambda_1$
and $\Lambda_2$ of a unitary operator
 $U\in U(4)$ and the structure of the spectrum of $U$:
a) generic case; b) double degeneracy; c) both eigenvalues doubly
degenerated; d) triple degeneracy.} \label{fig2}
\end{center}
    \end{figure}

The construction of rank-2 projections used in the previous proof
is typical in a certain sense. Ostensibly this follows because, as
proved above, these sets have no topological interior. They are
either finite discrete sets, or line segments. We shall provide
full details on this point in the next section. (In fact, this
result together with the discussion of the subsequent section can
be easily adjusted to verify the higher-rank numerical range
conjecture from \cite{CKZ05a} for arbitrary normal operators on
4-dimensional Hilbert space.) Of course, a further extension of
Theorem~\ref{evalues2qubit} to the case of arbitrary
$\Lambda_k(U)$ would greatly benefit from either the verification
of the conjecture, or, at the least, more substantive information
on these sets for arbitrary unitary operators.

We can establish the following as a consequence of the previous
proof.

\begin{theorem}\label{buc}
Let $\E = \{  V,  W \}$ be a BUC on a Hilbert space $\H$ with
$\dim\H\geq 4$. Then there are 2-dimensional code subspaces $\C$
of $\H$ such that $\C$ is correctable for $\E$.
\end{theorem}

{\noindent}{\it Proof.} As in Eq.~(\ref{simpleform}), we may
assume the channel is of the form $\E = \{\one,U\}$. Thus, the
theorem will be proved if we can show that $\Lambda_2(U)$ is
non-empty for an arbitrary unitary operator $U$.

In the case $\H = \bbC^4$, this result follows directly from
case~$(ii)$ of Theorem~\ref{evalues2qubit}. Indeed, observe that
$\Lambda_2(U)$ is shown to be non-empty in each of the subcases of
$(ii)$. The analysis of this case may be adapted to the case of
arbitrary $\H$. To see this, note first that if there is
degeneracy in the spectrum of $U$, say the eigenvalue $\lambda =
z_j$ has multiplicity at least two, then we can simply choose a
rank-2 sub-projection of the eigen-projection for $z_j$ to show
that $\Lambda_2(U)$ is non-empty in this case.

On the other hand, if there is no degeneracy in the spectrum of
$U$, then we may find four distinct eigenvalues for $U$ lying on
the unit circle, $z_j = e^{i\theta_j}$, $j=1,2,3,4$, ordered so
that $0\leq \theta_1 < \theta_2 < \theta_3 < \theta_4 < 2\pi$. As
in the previous proof, we can now use the constructive
``eigenvalue pairing'' approach to show that the intersection
point of the line through $z_1$ and $z_3$, with the line through
$z_2$ and $z_4$, belongs to the set $\Lambda_2(U)$. This completes
the proof.   \hfill$\square$

Let us note that Theorem~\ref{evalues2qubit} can also be used in a
negative fashion. For instance, consider the class of two-qubit
randomized unitary channels with at least three distinct unitary
error operators. We can conclude from Theorem~\ref{evalues2qubit}
that the set of such channels for which there exist correctable
codes forms a set of measure zero within the set of all such
channels. (Contrast this with the case of BUCs in
Theorem~\ref{buc}.) Indeed, if $\E = \{\one,U,V\}$ is a two-qubit
channel with $U\neq V$, and neither equal to a scalar multiple of
$\one$, then there are three pertinent equations to solve of the
form given in Eq.~(\ref{unitaryeqn}); namely, for the unitaries
$U$, $V$, and $U^\dagger V$. But we know that in the generic
non-degenerate case, the set $\Lambda_2$ will be a singleton, and
that there will almost never be a projection of rank at least two
that simultaneously solves Eq.~(\ref{unitaryeqn}) for all three
operators.

\vspace{0.1in}

\noindent{\it Correctable Codes for the Two-Qubit Case.} --- We
first discuss a simple example to illustrate the connection with
the stabilizer formalism. Let $\E=\{\one_4,ZZ\}$ be the error
model on $\H = \bbC^2\otimes\bbC^2$ for a channel that leaves
states alone with some probability, and applies the Pauli phase
flip operator $Z =\kb{0}{0} - \kb{1}{1}$ on each qubit with
another probability. We have chosen this particular example to
simplify the presentation, but a similar analysis applies to any
of the two-qubit Pauli-error models.

The stabilizer formalism applied to the error model $\E =
\{\one_4,ZZ\}$ obtains codes by considering the joint eigenspace
structure for the error operators $\{\one_4,ZZ\}$, and hence just
$ZZ = Z\otimes Z$ in this case. Through this approach one can find
the rank-2 projections
\begin{eqnarray}
P_1 &=& \kb{00}{00} + \kb{11}{11} \\
P_{-1} &=& \kb{01}{01} + \kb{10}{10},
\end{eqnarray}
as correctable codes for $\E=\{\one_4,ZZ\}$. Indeed,
Eq.~(\ref{correctcondition}) is satisfied since $P_{\pm 1}ZZ
P_{\pm 1} = ZZ P_{\pm 1} = \pm P_{\pm 1}$.

On the other hand, the higher-rank numerical range approach
captures these projections, and more. Specifically, as $ZZ$ is
Hermitian, Theorem~\ref{Hermitiancase} can be applied to obtain
$\Lambda_2(ZZ) = [-1,1]$. Thus, there is a family of rank-2
projections, each of which yields a correctable qubit code for $\E
= \{\one_4,ZZ\}$, for every element $\lambda\in [-1,1]$. For
instance, given $a,b,c,d\geq 0$ such that $a+b=1=c+d$, we can
define
\begin{eqnarray}\label{zzprojns}
\ket{\psi_1} &=& \sqrt{a} \ket{00} + \sqrt{b} \ket{01}  \\
\label{zzprojns1} \ket{\psi_2} &=& \sqrt{c} \ket{11}  + \sqrt{d}
\ket{10} ,
\end{eqnarray}
and define $P = \kb{\psi_1}{\psi_1} + \kb{\psi_2}{\psi_2}$. Then
$PZZP=\lambda P$ (though note it is not true in general that $PZZP
= ZZP$), and it follows that $P\H =
\spn\{\ket{\psi_1},\ket{\psi_2}\}$ is correctable for
$\E=\{\one_4,ZZ\}$. Notice that $P_1$ is captured by $b=d=0$ and
$P_{-1}$ by $a=c=0$. However, the cases $a\neq b$ and $c\neq d$
are not obtained as stabilizers of two-qubit Pauli-error models.

\begin{remark}
{\rm There are more projections that yield correctable codes for
$\E$, and we shall indicate this in the context of the next
example. We first wish to emphasize a basic point exposed by this
example. The operator equation $PAP=\lambda P$ does not
necessarily imply $AP = \lambda P$; i.e., if the compression of an
operator $A$ by a projection $P$ is a scalar multiple of $P$, then
not necessarily is the restriction of $A$ also a scalar multiple
of $P$. In terms of the block matrix decomposition of $A$ with
respect to $\H = P\H \oplus P^\perp \H$, these two statements can
be phrased visually as follows:
\begin{eqnarray}\label{compression}
PAP = \lambda P \qiff A = \left( \begin{matrix} \lambda\one_P &
\ast \\ \ast & \ast \end{matrix} \right)
\end{eqnarray}
\begin{eqnarray}
AP = \lambda P \qiff A = \left( \begin{matrix} \lambda\one_P &
\ast \\ 0 & \ast \end{matrix} \right).
\end{eqnarray}
It is evident that Eq~(\ref{compression}) allows for more
possibilities.

To be more precise, given an error model $\E = \{E_a\}$, one way
to seek correctable codes for $\E$ is to consider the joint
eigenspaces for the operators $E_a$ (as is done in the stabilizer
formalism applied for Pauli errors); i.e., projections $P$ such
that $E_a P = \lambda_a P$ for some scalar $\lambda_a$, for all
$a$. (Recall also that by Proposition~\ref{Hermitiantreduction},
the $E_a$ may be assumed to be Hermitian.) Then necessarily
Eq.~(\ref{correctcondition}) is satisfied for all $a,b$. However,
$PE_a^\dagger E_bP = \lambda_{ab} P$ does not in general imply
that $P$ is a joint eigen-projection for $E_a$ and $E_b$. Thus,
the previous example, and those that follow, are illustrative of a
more general phenomena. Namely, to capture all possible
correctable codes, we need to consider ``compressions'' of
operators instead of restricting ourselves to ``restrictions''. }
\end{remark}

Let us discuss, in more detail than the previous example, the
correctable code structure for the Pauli-error model given by $\E
= \{\one_4,Z_1\}$ on $\H = \bbC^2\otimes\bbC^2$, where
\[
Z_1 = Z\otimes\one_2 = \kb{00}{00} + \kb{01}{01} - \kb{10}{10} -
\kb{11}{11}.
\]
Here the structure of $\Lambda_2(Z_1) = [-1,1]$ determines the
code structure for $\E$. Let us consider the case $\lambda =
0\in\Lambda_2(Z_1)$. Let $P_+$ (respectively $P_-$) be the
projection onto the eigenspace $\V_+ = \spn\{ \ket{00},\ket{01}\}$
(respectively $\V_- = \spn\{ \ket{10},\ket{11}\}$). Then a typical
rank-2 code for $\E$ is given as follows. Let $\{
\ket{\psi_{\pm}}, \ket{\phi_{\pm}} \}$ be an orthonormal pair of
vectors in $\V_{\pm}$. Define
\begin{eqnarray}\label{Cform}
\C = \spn\{ \ket{\psi_+} + \ket{\psi_-}, \ket{\phi_+} +
\ket{\phi_-} \}.
\end{eqnarray}
One can verify that $PZ_1P = 0 = 0P$. For instance, if
$\ket{\psi}=\ket{\psi_+} + \ket{\psi_-}$, then
\begin{eqnarray}
\bra{Z_1\psi}\ket{\psi} = \bra{\psi_+}\ket{\psi_+} -
\bra{\psi_-}\ket{\psi_-} = 0.
\end{eqnarray}

On the other hand, suppose $\C$ is a two-dimensional subspace,
with projection $P$, such that $PZ_1P=0$. Then it follows that $\C
= P\H = P_+ P\H \oplus P_- P\H$, and that each of the subspaces
$P_{\pm}P\H$ is two-dimensional. Hence, as two-dimensional
subspaces of $\V_\pm$, we have $P_\pm P\H = \V_\pm$. Thus, $\C$
necessarily has the form given in Eq.~(\ref{Cform}). (See
\cite{CKZ05a} for a more complete discussion on this point in the
arbitrary Hermitian case.)

We finish by characterizing the correctable codes for a generic
BUC on two-qubit Hilbert space. As above, for error correction
purposes, we may assume such a channel has the form $\E = \{
\one_4,U\}$, where $U$ is a unitary on $\H=\bbC^4$. In the generic
case, the spectrum of $U$ will be non-degenerate, and so we can
assume the eigenvalues for $U$ are four distinct complex numbers
$\{z_1,z_2,z_3,z_4\}$, ordered so that $0\leq \arg z_1 < \ldots <
\arg z_4 < 2\pi$. Let $\ket{\psi_j}$ be corresponding
eigenvectors; $U\ket{\psi_j} = z_j \ket{\psi_j}$.

By Theorem~\ref{evalues2qubit}, the correctable qubit codes for
$\E$ correspond to the projections $P$ that satisfy
Eq.~(\ref{unitaryeqn}) for the unique $\lambda$ that belongs to
$\Lambda_2(U)$. Without loss of generality, we shall assume
$\lambda = 0$. (The reader will notice that the following argument
applies to normal operators. Hence, if $\Lambda_2(U)=\{\lambda\}$,
then we could replace $U$ by the normal operator
$U-\lambda\one_4$.)

Given that $\Lambda_2(U) = \{0\}$, we claim that
\begin{eqnarray}\label{2qubit0}
P = \kb{\xi_1}{\xi_1} + \kb{\xi_2}{\xi_2}
\end{eqnarray}
is a rank-2 projection such that
\begin{eqnarray}\label{compression0}
PUP = 0
\end{eqnarray}
if and only if there are angles
$\alpha_k,\beta_k,\gamma_k,\theta_{kj}$, for $k=1,2$ and
$j=2,3,4$, such that
\begin{eqnarray}\label{2qubit1}
\ket{\xi_k} = a_k \ket{\psi_1} +  b_k \ket{\psi_2} + c_k
\ket{\psi_3} + d_k \ket{\psi_4},
\end{eqnarray}
where
\begin{eqnarray}\label{2qubit2}
\left\{ \begin{array}{rcl} a_k &=& \cos\alpha_k \cos\beta_k \\ b_k
&=& e^{i\theta_{k2}}\sin\alpha_k \cos\gamma_k \\ c_k &=&
e^{i\theta_{k3}}\cos\alpha_k \sin\beta_k \\ d_k &=&
e^{i\theta_{k4}}\sin\alpha_k \sin\gamma_k \end{array}\right.,
\end{eqnarray}
and the following equations are satisfied;
\begin{eqnarray}\label{2qubit3}
\left\{ \begin{array}{rcl} \cos^2\beta_k z_1 +  \sin^2\beta_k z_3 &=& 0 \\
\cos^2\gamma_k z_2 +  \sin^2\gamma_k z_4 &=& 0 \end{array}\right.,
\end{eqnarray}
and
\begin{eqnarray}\label{2qubit4}
\left\{ \begin{array}{rcl} a_1a_2z_1 + b_1 \overline{b_2} z_2 +
c_1 \overline{c_2} z_3 + d_1 \overline{d_2} z_4  &=& 0 \\
a_1a_2z_1 +  \overline{b_1} b_2 z_2 + \overline{c_1}c_2 z_3 +
\overline{d_1}d_2 z_4  &=& 0 \end{array}\right..
\end{eqnarray}

To verify the sufficiency of these constraints for
Eq.~(\ref{compression0}), it must be shown that
$\bra{U\xi_k}\ket{\xi_l} = 0$ for $k,l =1,2$ when $\xi_1$ and
$\xi_2$ are defined as in Eq.~(\ref{2qubit1}) and the subsequent
equations. This follows from Eqs.~(\ref{2qubit3},\ref{2qubit4}),
and we leave this computation to the interested reader. Note the
special case of this construction in which the vectors are
obtained from the eigenvalue-pairing construction as in
Eq.~(\ref{rank2vectors}). (In that case, Eq.~(\ref{2qubit4}) is
trivially satisfied.)

On the other hand, for necessity, suppose $P$ and $\xi_1,\xi_2$
are given as in Eq.~(\ref{2qubit0}) and Eq.~(\ref{2qubit1}), and
that $PUP=0$. Then Eq.~(\ref{2qubit4}) comes as a direct
consequence of the identity $\bra{U\xi_1}\ket{\xi_2} = 0 =
\bra{U\xi_2}\ket{\xi_1}$. Further, for $k=1,2$ we have
\begin{eqnarray}
0 &=& \bra{U\xi_k}\ket{\xi_k} \\ &=& |a_k|^2z_1 + |b_k|^2z_2
+|c_k|^2z_3 +|d_k|^2z_4 .
\end{eqnarray}
In particular, this implies that
\begin{eqnarray}
|a_k|^2z_1 + |c_k|^2z_3   = -|b_k|^2z_2 -|d_k|^2z_4 =0,
\end{eqnarray}
as this equation describes the intersection point ($\lambda = 0$)
of the line through $z_1,z_3$ with the line through $z_2,z_4$.
This yields Eq.~(\ref{2qubit3}), and for succinctness we shall
leave the verification of the specific form of the scalars
$a_k,b_k,c_k,d_k$ given by Eq.~(\ref{2qubit2}) to the reader.

Let us briefly summarize how one can find all correctable codes
for any given bi-unitary channel $\E = \{V,W\}$ on $\H = \bbC^4$:
\begin{itemize}
\item[$(i)$] Compute the set of compression-values
$\Lambda_2(V^\dagger W)$ via Theorem~\ref{evalues2qubit}.
\item[$(ii)$] For each compression-value $\lambda$ from $(i)$, the
family of projections $P$ that satisfy $PV^\dagger WP = \lambda P$
may be obtained as in the discussion of this section.
\item[$(iii)$] The subspaces corresponding to ranges of
projections from $(ii)$ are precisely the correctable codes for
$\E = \{V,W\}$.
\end{itemize}




\vspace{0.1in}

\noindent{\it Conclusion.} --- We have solved the fundamental
error correction problem in quantum computing for bi-unitary
channels on two-qubit Hilbert space. This was accomplished through
an application of a new tool -- the ``higher-rank numerical
range'' -- that has been recently developed to solve algebraic
compression problems. We have shown that, in the case of
Pauli-error models, this approach captures codes not obtained
through the stabilizer formalism for Pauli matrices. We also
discussed further applications to more general quantum channels on
larger Hilbert spaces. As an example, we constructed qubit codes
for bi-unitary channels on arbitrary Hilbert spaces. (Compression
error correcting codes for bi-unitary channels on arbitrary
Hilbert spaces will be analyzed in a forthcoming article
\cite{CHKZ06}.) We also discussed how this approach can be used to
establish negative results.

To apply the information we have derived for the higher-rank
numerical ranges in the Hermitian case more generally, a better
understanding of the intersections of projections in
Eq.~(\ref{correctcondition}) is required. As another avenue to
more general applications, a complete understanding of the
higher-rank numerical range for the case of normal operators, or
even unitary operators, could help in special cases such as the
class of randomized unitary channels. We have not considered
possible implications of the higher-rank numerical ranges to
problems in the new protocol for error correction called
``operator quantum error correction'' \cite{KLP05,KLPL05}. In this
scheme, a characterization of correction has been obtained
\cite{KLPL05,NP05}  that generalizes Eq.~(\ref{correctcondition}),
and it may be possible to apply these tools to that more general
setting.

\vspace{0.1in}

\noindent{\it Acknowledgements.} We would like to thank Daniel
Gottesman and Raymond Laflamme for helpful discussions. We are
also grateful to other colleagues at IQC, PI, and UofG for
stimulating conversations. M.D.C. and D.W.K. were partially
supported by NSERC. K.~{\.Z}. is thankful to Perimeter Institute
for creating optimal working conditions in Waterloo and
acknowledges a partial support by the grant number
PBZ-MIN-008/P03/2003 of Polish Ministry of Science and Information
Technology.


\begin{thebibliography}{30}
\expandafter\ifx\csname
natexlab\endcsname\relax\def\natexlab#1{#1}\fi
\expandafter\ifx\csname bibnamefont\endcsname\relax
  \def\bibnamefont#1{#1}\fi
\expandafter\ifx\csname bibfnamefont\endcsname\relax
  \def\bibfnamefont#1{#1}\fi
\expandafter\ifx\csname citenamefont\endcsname\relax
  \def\citenamefont#1{#1}\fi
\expandafter\ifx\csname url\endcsname\relax
  \def\url#1{\texttt{#1}}\fi
\expandafter\ifx\csname
urlprefix\endcsname\relax\def\urlprefix{URL }\fi
\providecommand{\bibinfo}[2]{#2}
\providecommand{\eprint}[2][]{\url{#2}}





\bibitem[{\citenamefont{Shor}(1995)}]{Sho95a}
\bibinfo{author}{\bibfnamefont{P.~W.} \bibnamefont{Shor}},
  \bibinfo{journal}{Phys. Rev. A} \textbf{\bibinfo{volume}{52}},
  \bibinfo{pages}{R2493} (\bibinfo{year}{1995}).

\bibitem[{\citenamefont{Steane}(1996)}]{Ste96a}
\bibinfo{author}{\bibfnamefont{A.~M.} \bibnamefont{Steane}},
  \bibinfo{journal}{Phys. Rev. Lett.} \textbf{\bibinfo{volume}{77}},
  \bibinfo{pages}{793} (\bibinfo{year}{1996}).

  \bibitem[{\citenamefont{Bennett et~al.}(1996)\citenamefont{Bennett,
DiVincenzo,
  Smolin, and Wootters}}]{BDSW96a}
\bibinfo{author}{\bibfnamefont{C.~H.} \bibnamefont{Bennett}},
  \bibinfo{author}{\bibfnamefont{D.~P.} \bibnamefont{DiVincenzo}},
  \bibinfo{author}{\bibfnamefont{J.~A.} \bibnamefont{Smolin}},
  \bibnamefont{and} \bibinfo{author}{\bibfnamefont{W.~K.}
  \bibnamefont{Wootters}}, \bibinfo{journal}{Phys. Rev. A}
  \textbf{\bibinfo{volume}{54}}, \bibinfo{pages}{3824} (\bibinfo{year}{1996}).

\bibitem[{\citenamefont{Knill and Laflamme}(1997)}]{KL97a}
\bibinfo{author}{\bibfnamefont{E.}~\bibnamefont{Knill}} \bibnamefont{and}
  \bibinfo{author}{\bibfnamefont{R.}~\bibnamefont{Laflamme}},
  \bibinfo{journal}{Phys. Rev. {A}} \textbf{\bibinfo{volume}{55}},
  \bibinfo{pages}{900} (\bibinfo{year}{1997}).

\bibitem[{\citenamefont{Gottesman}(1996)}]{Got96}
\bibinfo{author}{\bibfnamefont{D.} \bibnamefont{Gottesman}},
  \bibinfo{journal}{Phys. Rev. A} \textbf{\bibinfo{volume}{54}}
 \bibinfo{pages}{1862} (\bibinfo{year}{1996}).

\bibitem[{\citenamefont{Gottesman}(1997)}]{Got97}
\bibinfo{author}{\bibfnamefont{D.} \bibnamefont{Gottesman}},
  \bibinfo{journal}{Ph.D. thesis, California Institute of Technology, Pasadena, CA}
  (\bibinfo{year}{1997}).

  \bibitem[{\citenamefont{Calderbank, et~al.}(1997)}]{CRSS97}
\bibinfo{author}{\bibfnamefont{A.~R.} \bibnamefont{Calderbank}},
  \bibinfo{author}{\bibfnamefont{E.~M.} \bibnamefont{Rains}},
  \bibinfo{author}{\bibfnamefont{P.~W.} \bibnamefont{Shor}},
  \bibnamefont{and} \bibinfo{author}{\bibfnamefont{N.~J.~A.}
  \bibnamefont{Sloane}}, \bibinfo{journal}{Phys. Rev. Lett.}
  \textbf{\bibinfo{volume}{78}}, \bibinfo{pages}{405} (\bibinfo{year}{1997}).

  \bibitem[{\citenamefont{Rains, et~al.}(1997)\citenamefont{ }}]{RHSS97}
\bibinfo{author}{\bibfnamefont{E.~M.} \bibnamefont{Rains}},
\bibinfo{author}{\bibfnamefont{R.~H.} \bibnamefont{Hardin}},
  \bibinfo{author}{\bibfnamefont{P.~W.} \bibnamefont{Shor}},
  \bibnamefont{and} \bibinfo{author}{\bibfnamefont{N.~J.~A.}
  \bibnamefont{Sloane}}, \bibinfo{journal}{Phys. Rev. Lett.}
  \textbf{\bibinfo{volume}{79}}, \bibinfo{pages}{953} (\bibinfo{year}{1997}).

\bibitem[{\citenamefont{Aharonov and Ben-Or}(1999)}]{ABO99}
\bibinfo{author}{\bibfnamefont{D.}~\bibnamefont{Aharonov}} \bibnamefont{and}
  \bibinfo{author}{\bibfnamefont{M.}~\bibnamefont{Ben-Or}},
  \bibinfo{journal}{arxiv.org/quant-ph/9906129}.

\bibitem[{\citenamefont{Rains}(1999)}]{Rai99b}
\bibinfo{author}{\bibfnamefont{E.~M.} \bibnamefont{Rains}},
  \bibinfo{journal}{IEEE Trans. Inf. Theory}
  \textbf{\bibinfo{volume}{45}},
 \bibinfo{pages}{2361} (\bibinfo{year}{1999}).

\bibitem[{\citenamefont{Nielsen and Chuang}(2000)}]{NC00}
\bibinfo{author}{\bibfnamefont{M.~A.}~\bibnamefont{Nielsen}} \bibnamefont{and}
  \bibinfo{author}{\bibfnamefont{I.~L.}~\bibnamefont{Chuang}},
  \bibinfo{journal}{Cambridge University Press,} (\bibinfo{year}{2000}).

\bibitem[{\citenamefont{Klappenecker and Roetteler}(2002)}]{KlRo02a}
\bibinfo{author}{\bibfnamefont{A.}~\bibnamefont{Klappenecker}} \bibnamefont{and}
  \bibinfo{author}{\bibfnamefont{M.}~\bibnamefont{Roetteler}},
  \bibinfo{journal}{IEEE Trans. Inf.
  Theory}\textbf{\bibinfo{volume}{48}},
 \bibinfo{pages}{2392} (\bibinfo{year}{2002}).

\bibitem[{\citenamefont{Pollatsek and Ruskai}(2004)}]{PoRu04}
\bibinfo{author}{\bibfnamefont{H.}~\bibnamefont{Pollatsek}} \bibnamefont{and}
  \bibinfo{author}{\bibfnamefont{M.~B.}~\bibnamefont{Ruskai}},
  \bibinfo{journal}{Lin. Alg.
  Appl.}\textbf{\bibinfo{volume}{392}},
 \bibinfo{pages}{255} (\bibinfo{year}{2004}).

\bibitem[{\citenamefont{Gottesman}(2005)}]{Got05}
\bibinfo{author}{\bibfnamefont{D.} \bibnamefont{Gottesman}},
  \bibinfo{journal}{arxiv.org/quant-ph/0507174}.

\bibitem{CKZ05a}
  \bibinfo{author}{\bibfnamefont{M.~D.}~\bibnamefont{Choi}},
\bibinfo{author}{\bibfnamefont{D.~W.}~\bibnamefont{Kribs}}, \bibnamefont{and}
  \bibinfo{author}{\bibfnamefont{K.}~\bibnamefont{{\.Z}yczkowski}},
 \bibinfo{journal}{Lin. Alg. Appl., to appear}.

\bibitem[{\citenamefont{Choi}(1975)}]{Cho75}
\bibinfo{author}{\bibfnamefont{M.~D.} \bibnamefont{Choi}},
  \bibinfo{journal}{Lin. Alg. Appl.}
  \textbf{\bibinfo{volume}{10}},
 \bibinfo{pages}{285-290} (\bibinfo{year}{1975}).

\bibitem[{\citenamefont{Kraus}(1971)}]{Kra71}
\bibinfo{author}{\bibfnamefont{K.} \bibnamefont{Kraus}},
  \bibinfo{journal}{Ann. Physics} \textbf{\bibinfo{volume}{64}},
 \bibinfo{pages}{311-335} (\bibinfo{year}{1971}).

\bibitem{Hal67}
\bibinfo{author}{\bibfnamefont{P.}~\bibnamefont{Halmos}},
   \bibinfo{journal}{D. Van Nostrand Company, Ltd., Toronto,} (\bibinfo{year}{1967}).

\bibitem{AL87}
\bibinfo{author}{\bibfnamefont{R.}~\bibnamefont{Alicki}}, \bibnamefont{and}
  \bibinfo{author}{\bibfnamefont{K.}~\bibnamefont{Lendi}},
  \bibinfo{journal}{Springer--Verlag, Berlin}, (\bibinfo{year}{1987}).

\bibitem{KLP05}
\bibinfo{author}{\bibfnamefont{D.~W.}~\bibnamefont{Kribs}},
  \bibinfo{author}{\bibfnamefont{R.}~\bibnamefont{Laflamme}} \bibnamefont{and}
  \bibinfo{author}{\bibfnamefont{D.}~\bibnamefont{Poulin}},
  \bibinfo{journal}{Phys. Rev. Lett.},\textbf{\bibinfo{volume}{94}},
  \bibinfo{pages}{180501} (\bibinfo{year}{2005}).

\bibitem{KLPL05}
\bibinfo{author}{\bibfnamefont{D.~W.}~\bibnamefont{Kribs}},
  \bibinfo{author}{\bibfnamefont{R.}~\bibnamefont{Laflamme}},
  \bibinfo{author}{\bibfnamefont{D.}~\bibnamefont{Poulin}} \bibnamefont{and}
  \bibinfo{author}{\bibfnamefont{M.}~\bibnamefont{Lesosky}},
 \bibinfo{journal}{arxiv.org/quant-ph/0504189}.

\bibitem{NP05}
\bibinfo{author}{\bibfnamefont{M.~A.}~\bibnamefont{Nielsen}} \bibnamefont{and}
  \bibinfo{author}{\bibfnamefont{D.}~\bibnamefont{Poulin}},
  \bibinfo{journal}{arxiv.org/quant-ph/0506069}.

\bibitem{CHKZ06}
  \bibinfo{author}{\bibfnamefont{M.~D.}~\bibnamefont{Choi}},
  \bibinfo{author}{\bibfnamefont{J.~A.}~\bibnamefont{Holbrook}},
\bibinfo{author}{\bibfnamefont{D.~W.}~\bibnamefont{Kribs}}, \bibnamefont{and}
  \bibinfo{author}{\bibfnamefont{K.}~\bibnamefont{{\.Z}yczkowski}},
 \bibinfo{journal}{in preparation}.


\end{thebibliography}


\end{document}